\documentclass[conference]{IEEEtran}
\usepackage{graphicx}
\usepackage{float} 
\usepackage{url} 
\usepackage{color}

  \usepackage{caption}
  \usepackage{subcaption}
  \usepackage{setspace}

\newenvironment{noindlist2}
 {\begin{list}{\labelitemi}{\leftmargin=1.1em \itemindent=-0.6em}}
 {\end{list}}

\ifCLASSINFOpdf
\else
\fi
\begin{document}
%
\title{MOSDEN: An Internet of Things Middleware for Resource Constrained Mobile Devices}

\author{\IEEEauthorblockN{Charith Perera, Prem Prakash Jayaraman,\\ Arkady Zaslavsky,  Dimitrios Georgakopoulos\\ }
\IEEEauthorblockA{CSIRO ICT Center, Canberra, ACT 2601, Australia\\
\{charith.perera, prem.jayaraman, arkady.zaslavsky, \\  dimitrios.georgakopoulos\}@csiro.au}
\and
\IEEEauthorblockN{Peter Christen}
\IEEEauthorblockA{Research School of Computer Science,\\ The Australian National University, \\Canberra, ACT 0200, Australia\\
peter.christen@anu.edu.au}
}


%


\maketitle

\begin{abstract}
The Internet of Things (IoT) is part of Future Internet and will comprise many billions of Internet Connected Objects (ICO) or \textit{`things'} where things can sense, communicate, compute and potentially actuate as well as have intelligence, multi-modal interfaces, physical/ virtual identities and attributes. Collecting data from these objects is an important task as it allows software systems to understand the environment better. Many different  hardware devices may involve in the process of collecting and uploading sensor data to the cloud where complex processing can occur. Further, we cannot expect all these objects to be connected to the computers due to technical and economical reasons. Therefore, we should be able to utilize resource constrained devices to collect data from these ICOs. On the other hand, it is critical to process the collected sensor data before sending them to the cloud to make sure the sustainability of the infrastructure due to energy constraints. This requires to move the sensor data processing tasks towards the resource constrained computational devices (e.g. mobile phones). In this paper, we propose Mobile Sensor Data Processing Engine (MOSDEN), an plug-in-based IoT middleware for mobile devices, that allows to collect and process sensor data  without programming efforts. Our architecture also supports sensing as a service model. We present the results of the evaluations that demonstrate its suitability towards real world deployments. Our proposed middleware is built on Android platform.

\end{abstract}


%
\IEEEpeerreviewmaketitle

\section{Introduction}
\label{sec:Introduction}

\begin{figure*}[t]
 \includegraphics[scale=.90]{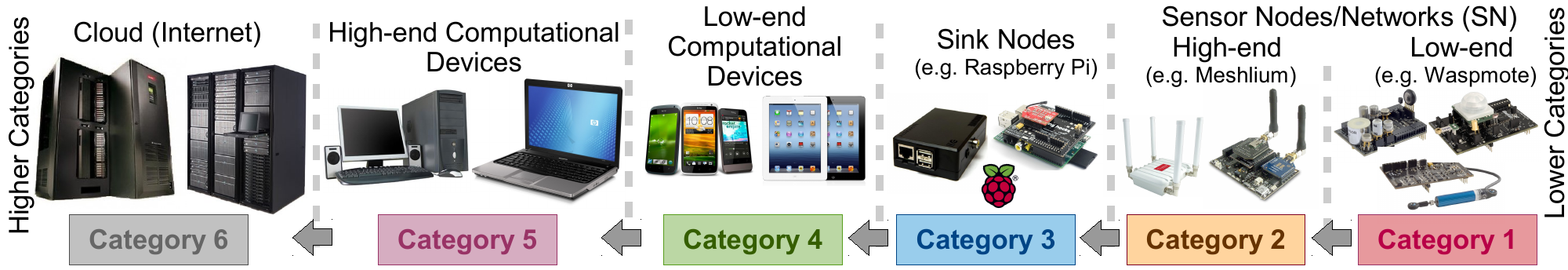}
 \caption{Categorization of IoT devices based on their computational capabilities. The devices belong to each category has different capabilities depending in term of processing, memory, and communication. They are also different in price where devices becomes more expensive when move to left. The computational capabilities are also get increased when move to left.}
 \label{Figure:Layered_Architecture}	
\vspace{-0.33cm}	
\end{figure*}

Internet of Things (IoT) \cite{P003} is a part of future Internet and ubiquitous computing. It envisions interactions between \textit{things\footnote{We use terms \textit{objects}, \textit{things}, \textit{smart objects}, \textit{devices}, \textit{nodes} to give the same meaning as they are frequently used in IoT related documentation interchangeably.}} that consists of sensors and actuators. As the price of sensors diminishes rapidly, we can soon expect to see very large numbers of \textit{things}. The vision of IoT is to allow \textit{`things'} to be connected anytime, anyplace, with anything and anyone, ideally using any path, any network and any service \cite{P029}. In order to realise this vision, we need a common operating platform namely middleware that is scalable and supports high level of interoperability. This platform enables sensor data collection, processing, and analysis. Efficient and feature rich IoT middeware platforms are key enablers of IoT paradigm. We are currently observing an emerging trend in middelware solutions that enable IoT \cite{P579}. However, most of the solutions are designed and developed to be used in the cloud environments where abundant resources are available. We believe that middleware solutions designed specifically for low powered resource constrained computation devices are critical in order realise the vision on IoT.

In this paper, we propose an IoT middleware solution that can work on resource constrained mobile devices allowing them to collect and process data from sensors easily. To achieve this, we extend existing middleware solution namely Global Sensor Network (GSN) \cite{P022} as well as propose new strategies that make our solution more scalable and user friendly. The contribution of this paper can be summarised as follows:

\begin{itemize}
\item We present the design and implementation details of our proposed middleware solution namely Mobile Sensor Data Processing Engine (MOSDEN). MOSDEN is designed to support sensing as a service model \cite{ZMP008} natively. Further, MOSDEN is a \textit{true zero programming} middleware where users do not need to write program code or any other specifications using declarative languages. Our solution also supports both push and pull data streaming mechanism as well as centralised and decentralised (e.g. peer-to-peer) data communication.

\item  We employ a plugin architecture, so developers can develop plugins allowing MOSDEN to communicate with for their sensor hardware. We also utilize the application markets that are built around android platform to efficiently share and distribute plugins.


\item We designed and developed MOSDEN in such a way that it is interoperable with other cloud-based middleware solutions such as GSN. Our pluggable architecture is scalable and promotes ease-of-use.


\item We present results of evaluating the performance of MOSDEN using devices with different capabilities and resource constraints in order to validate MOSDEN's scalability and suitability towards IoT domain.
\end{itemize}

The rest of this paper is structured as follows. Section \ref{sec:Background_and_Motivation} presents the background and the motivation. We set the background in two different perspectives: Internet of Things architecture and sensing as a service model. In Section \ref{sec:Research_Problem}, we defines the research challenges that we have addressed in this paper. Subsequently, Section \ref{sec:Architectural_Design} discusses the architectural design in details. Implementation details are presented in Section \ref{sec:Implementation}. Section \ref{sec:Evaluation} presents the evaluation of our IoT middleware using three mobile devices with different resource limitations. We also discuss the results in detail followed by an analysis of lessons learnt. Related literature is reviewed in Section \ref{sec:Related_Work} under three broad themes. Section \ref{sec:Conclusion} concludes the paper by highlighting future work.

\section{Background and Motivation}
\label{sec:Background_and_Motivation}
In this section, we briefly discuss the background and our motivation behind this work. The discussion is mainly structured under two main areas. First, we explain the hardware infrastructure that IoT is intended to use. Second, we explain the sensing as a service model and how it can fuels the adaptation of IoT. Our work is also stimulated and motivated by the statistics and prediction related to Internet of Things. Some of the details are discussed in \cite{ZMP003}.

\subsection{Internet of Things Architecture}
\label{sec:BM:Layered_Internet_of_Things_Architecture}

Even though IoT envisions billions of \textit{`things'} to be connected to the Internet, it is not possible and practical to connect all of them to the Internet directly. This is mainly due to resource constraints (e.g. network communication capabilities and energy limitations). Connecting directly to the internet is expensive in term of computation, bandwidth usage, and hardware cost point of view. Enabling persistent Internet access is challenging and also negatively impacts on miniaturization and energy consumption of the sensors. Due to such difficulties, IoT solutions need to utilize different types of devices with different resource limitations and capabilities. In Figure \ref{Figure:Layered_Architecture}, we broadly categorise these devices into 6 categories (also called level or layer).

The computational and connectivity capabilities increase as we move from right to left. Similarly, devices are also getting expensive and larger in form-factor when moving to the left. We believe that an ideal IoT middleware solution should be able to take advantage and adapt to  these different types of devices in order to make the solution more efficient and effective. One of the most critical decision that needs to be taken in the domain of IoT is  \textit{`where and `when'} to process the collected data. It is clear that no single solution would fit  every situation. Though there are many factors need to be considered, energy consumption for data processing and network communication are among the most important factors. If we denote the energy requirement for data processing as $E^{\alpha}$ and energy requirement for data communication over network as $E^{\beta}$, the following rule can be used to determine whether to process data in the current layer or send them  to a higher layer.

\begin{noindlist2}
\item IF ($E^{\alpha} < E^{\beta} $ ) THEN process locally ELSE send to a node with higher capability
\end{noindlist2}

Processing data in any device locally before sending them to the higher layers is important in terms of saving energy. However, the type of processing that needs to be performed at the each device is a difficult choice. Expensive processing can drain the battery quickly. In contrast, sending data frequently can also drain the battery quickly due to usage of communication radio \cite{ZMP001}. One of our motivations in this work is to address this problem. MOSDEN performs data processing and analytic before transmitting them over a network. More importantly, our proposed middleware platform can be installed on devices that belongs to lower level categories which have  resource limitations similar to mobile phones or Raspberry Pi\footnote{The Raspberry Pi is a credit-card-sized single-board computer developed in the UK by the Raspberry Pi Foundation with the intention of promoting the teaching of basic computer science in schools. It has maximum of 512MB memory, up to 1 GHz CPU. One unit cost around \$25. http://www.raspberrypi.org/}. For prototype implementation and evaluation, we use mobile phones. However, we believe, more cheaper devices with similar resource limitations will be available in the market overtime. MOSDEN can process sensor data based on SQL-like queries such as average which reduces the network communication due to sensor data fusion. The processing capabilities are discussed further in upcoming sections. Fifty different sensor application domains are explained in \cite{P416}. MOSDEN can be used in all these applications to improve the long-term sustainability of the IoT infrastructures by using available energy optimally and reduce unnecessary data communication. Specially, in outdoor sensing applications, it is troublesome and labour intensive to recharge the batteries of the sensing devices frequently.

\subsection{Sensing as a service Model}
\label{sec:BM:Sensing_as_a_service}

\begin{figure}[t]
 \centering
 \includegraphics[scale=0.78]{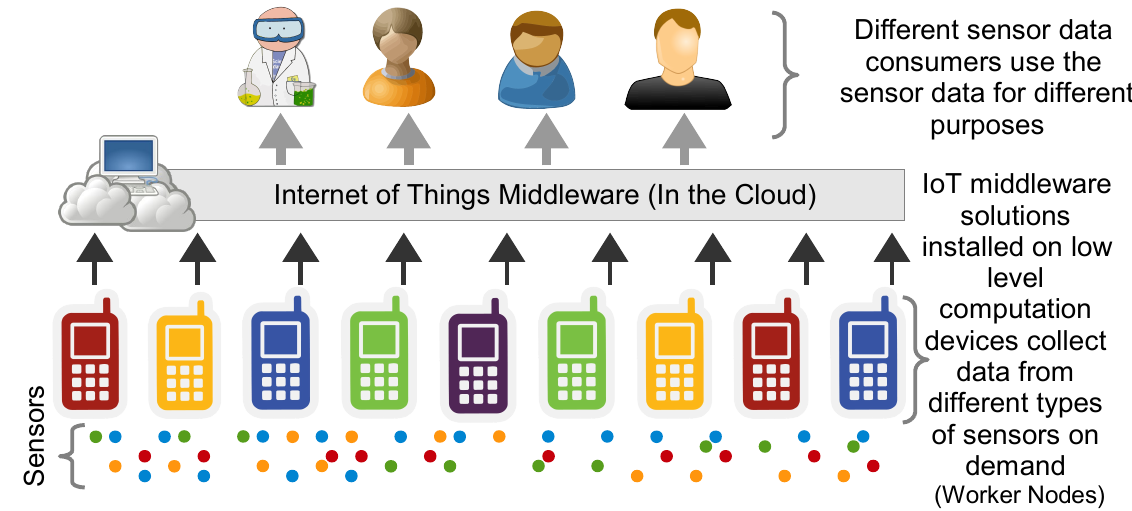}
\vspace{-0.43cm}	
 \caption{MOSDEN supports sensing as a service Model. Sensors that do not have long-range network communication capabilities connect to MOSDEN instances. Then, MOSDEN process the data and transmit them to the cloud selectively,}
 \label{Figure:Sensing_as_a_service}	
\vspace{-0.43cm}	
\end{figure}

This model provides sensors data to the users / consumer (i.e. anyone need access to sensor data) on-demand \cite{ZMP003}. Sensing as a service model does not collect sensor data from all the available sensors at all times. IoT middleware platforms that supports sensing as a service do keep track of the individual sensors, their availability, and capabilities. However, they do not collect sensor data unless a consumer makes a request. Our solution, MOSDEN, supports sensing as a service model. Specifically, MOSDEN provides easy way to retrieve data from sensors. MOSDEN also collect information about each sensor sends them to the cloud-based IoT middelware (e.g. GSN \cite{P022}). Cloud IoT middleware maintains a registry of sensors which includes their availabilities and capabilities. Cloud IoT middleware retrieves this information though multiple MOSDEN instances. Once the  Cloud IoT middleware receives a request from a consumer, it searches the required sensors and composes a request query. It then registers the request with the MOSDEN instance connected to the  sensor.

Then, MOSDEN sends data to Cloud IoT middleware until the request expires. Figure \ref{Figure:Sensing_as_a_service} illustrates the typical architecture of a sensing as a service model and the role of MOSDEN intends to play. More importantly, MOSDEN performs data processing before send the data to the server. For example, instead of sending data every 2 seconds, MOSDEN may locally process, store the data and send the data to the cloud once a minute by averaging values\footnote{The sampling rate and the data processing operation that may exactly use in each situation depends on the user requirement}. As another example, MOSDEN may collect all sensor data for a minute and send them to GSN at once. Such approach can save significant amount of energy due to reduction of network operations (e.g. opening and closing communication radios are energy expensive operations).

 Additionally, MOSDEN locally keeps track of the availabilities and capabilities of the sensors attached to each instance which makes it is easy and efficient for the cloud IoT middleware to stay up to date. Further, different MOSDEN instances connected to  cloud IoT middleware are managed using a publish / subscribe model \cite{P634}.

\section{Research Problem}
\label{sec:Research_Problem}

We address several research problems in this work. Our focus areas are energy efficient and effective data processing and network communication, cost efficient infrastructure support for large scale IoT deployment, and usability in connecting / configuring sensors. In the earlier section, we highlighted the importance of addressing the above mentioned research challenges: (1) the importance of  processing data locally before transmitting to the cloud, (2) the importance of utilizing devices with different computational capabilities and price tags, and (3) the importance of  providing efficient and easy way to connect sensors to low level computational devices (devices belongs to category 3 and 4 in Figure \ref{Figure:Layered_Architecture}). 

There are several commercial solutions\footnote{TWINE (supermechanical.com), Ninja Blocks (ninjablocks.com), and Smart Things (smartthings.com)} that have been proposed in order to address some of the above mentioned challenges. However, these solution have several weakness. The following brief analysis helps to identify those weaknesses as well as to identify the ideal design requirements of an IoT middleware that needs to be installed on resource constrained devices. \textcolor{black}{Though some of the hardware components are open sourced, software systems remain closed source which makes it hard to extend and interoperate. Further, these solutions have their own hardware devices that performs tasks similar to MOSDEN. However, these devices are custom built. We believe, utilizing commonly available devices such as mobile phones, makes it easy to adopt due to the fact the most of the people are familiar with mobile phones and know how to operate them in comparison to custom build proprietary devices. Another major drawback is inability for devices to interoperate with solutions provided by  different vendors. For example, a sensor designed to be used by one solution cannot be connected to the software system of another solution.}. Hence, our proposed middleware aims to be vendor agnostic.

\section{MOSDEN: Architectural Design}
\label{sec:Architectural_Design}
In this section, we explain the design decisions in details. First, we present the reasons behind introducing a plugin architecture. Secondly, we explain the complete MOSDEN architecture. Thirdly, we explain how MOSDEN interacts with its peers as well as cloud companions. Finally, we explain how distributed processing performed collectively by cloud companions (i.e. GSN instances) and MOSDEN instances.

\begin{figure*}[t]
 \centering
 \includegraphics[scale=1.1]{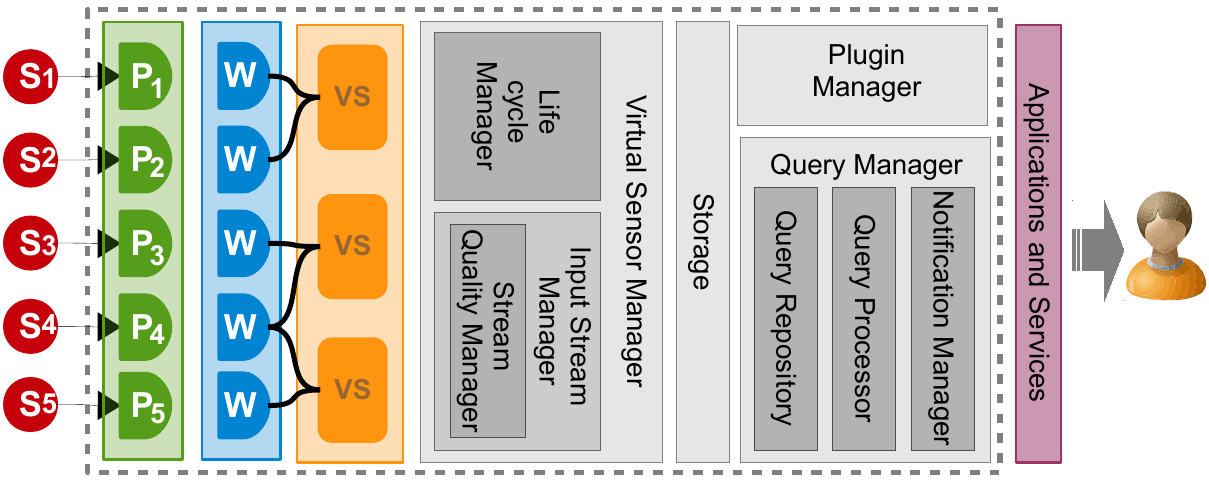}
 \vspace{-0.21cm}	
 \caption{The architectural design of the MOSDEN. Legend: Sensor (S), Plugin (P), Wrapper (W), Virtual Sensor (VS). Plugins communicates with the sensors and retrieve data. Each plugin should be compatible with the sensor it wants to communicates with. Plugins compatible with different sensors can be downloaded from Google Play}
 \label{Figure:Architecture}	
\vspace{-0.53cm}	
\end{figure*}

\subsection{Plugin Architecture}
\label{sec:A:Plugin_Architecture}
In MOSDEN, we employed a plugin architecture \cite{P632} in order to support three main requirements: scalability, usability, and community based development. A plugin  is a independent software component that adds a specific feature to an existing software application. In MOSDEN, each plugin translates generic communication messages to sensor specific commands in order to enable communication between MOSDEN and a specific sensor. When an application supports plugins, it enables customization. Further, MOSDEN plugins can be installed and configured at run time.

\textbf{Scalability:} Due to plugin approach, MOSDEN can virtually support any sensor in the world. Anyone can develop plugins that allow MOSDEN to communicate with given sensors. Further, plugins consumes very small amount of storage space (e.g. 25KB). Therefore, large number of plugins can be stored even in a resource limited mobile device. Furthermore,  MOSDEN automatically removes unused plugins when the memory is running low. New plugins can be downloaded through application stores such as \textit{Google Play} or directly as .apk files. Separation of plugins from the main MOSDEN application, helps to reduce the size of the application and also promotes plug-n-play. Practically, at a given point of time, only small number of plugins need to be installed in order to facilitate sensor communication though thousands of plugins would be available on applications stores. Finally, the plugin architecture allows us to improve MOSDEN in the future, specially in the directions of automated sensor discovery and plugin installation based on context information.

\textbf{Usability:} MOSDEN is convenient to use as it allows to collect data from sensors without programming efforts. Users are only required to download the matching plugin from an application store. Due to standardise plugin structure, MOSDEN knows how to communicate with each plugin. For the user, all the technical complexities and details are hidden and happen autonomously behind the scene.

\textbf{Community-based Development:} Plugin architecture allows us to engage with developer communities and support variety of different sensors through community-based development. Our software are expected to release as free and open source software in the future. We provide the main MOSDEN application as well as the standard interfaces where developers can  use to start develop their own plugins to support different sensors. We provide a sample plugin source code where developers only need to add their code according to the guidelines provided. Plugin model support to increasingly enable the number of sensors supported by MOSDEN. Plugins for MOSDEN can be downloaded via applications stores such as Google play.




\subsection{General Architecture}
\label{sec:A:General_Architecture}

The architecture of MOSDEN is presented in Figure \ref{Figure:Architecture}. MOSDEN architecture is based on the GSN architecture \cite{P022}. Additionally, we made several changes to the architecture in order to improve the efficiency as well as scalability. The major change is that we added a plugin manager and a plugin layer to support and manipulate plugins. GSN requires different wrappers to connect to different sensors. We eliminate this requirement and instead developed a single generic wrapper to handle the communication. In MOSDEN, wrappers do not directly communicate with sensors. Instead, the generic wrapper communicates with plugins and plugin communicates with the sensors (i.e. wrapper $\rightarrow$ plugins (P$_{i}$)$\rightarrow$ Sensor (S$_{i}$)). Due to the introduction of a generic wrapper, manual re-compilation of MOSDEN is not required when new sensors are added. Our newly added plugin manager component communicates with the cloud based GSN instances as well as MOSDEN peer instances and share the information about the sensors connected to them. All the other architectural components behave as same as in the GSN middleware \cite{P022}.


\subsection{Interaction with the Cloud and Peers}
\label{sec:A:Interaction_with_the_Cloud_and_Peers}

\begin{figure}[b]
 \centering
 \includegraphics[scale=0.40]{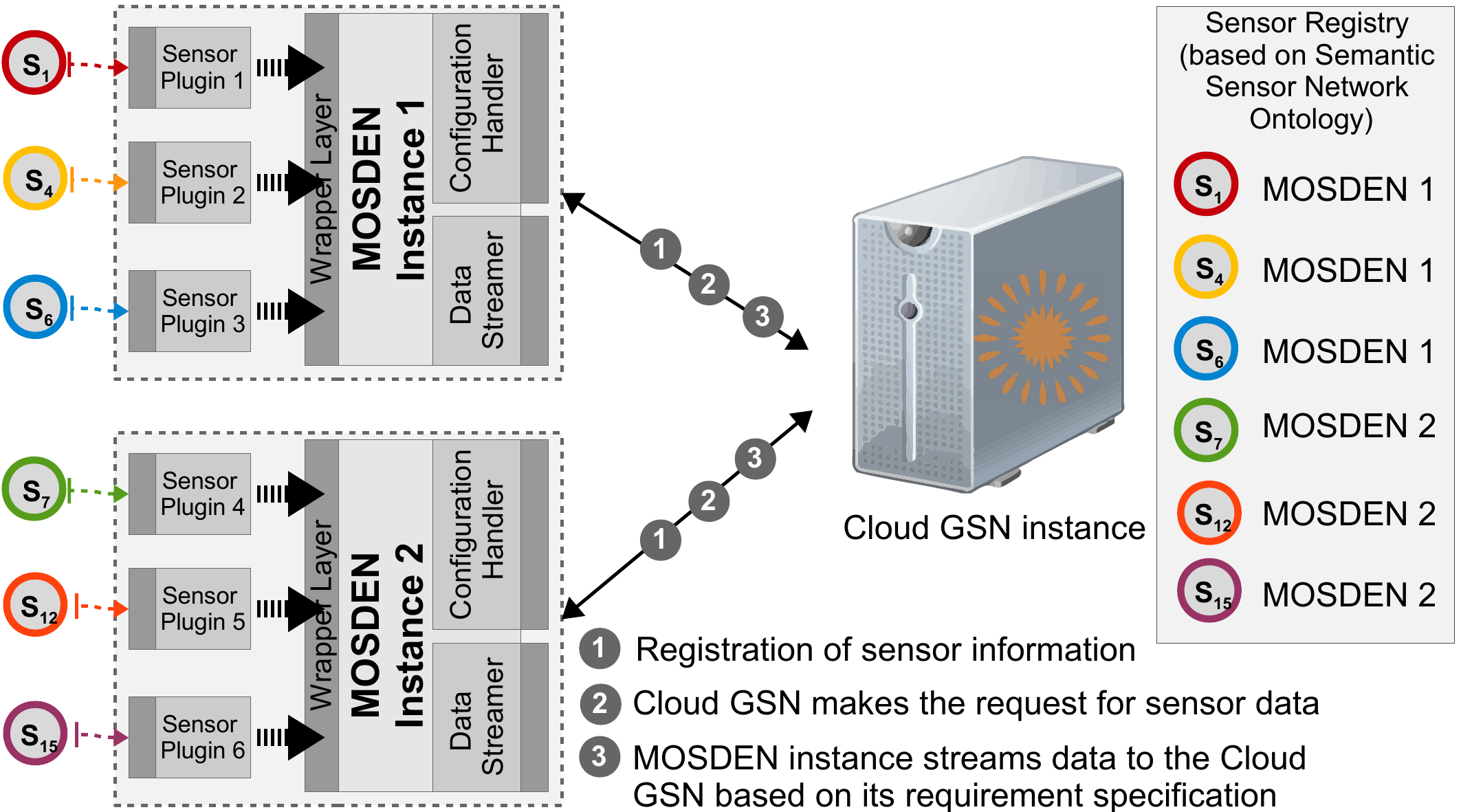}
\vspace{-2pt}	
 \caption{Interactions between MOSDEN and Cloud GSN}
 \label{Figure:System_Overview}	
\end{figure}

MOSDEN is design to be used as part of the sensing as a service model. On the other hand, due to that fact that our code is based on GSN middleware, MOSDEN is 100\% compatible with GSN. This means communication between GSN instances and MOSDEN instance can be performed natively without any additional effort. Further, MOSDEN is a part of our overall vision of providing middleware support across different categories of devices as depicted in Figure \ref{Figure:Layered_Architecture}. The typical interactions between GSN cloud instance and MOSDEN instances are illustrated in Figure \ref{Figure:System_Overview}. There are three main interaction that are frequently performed between MOSDEN instances and a GSN cloud instance. During our work, we also extended the cloud GSN architecture in order to support these interactions. When MOSDEN instance detect a new sensor connected to it through a plugin, it retrieves additional context information about the sensor (e.g. type of the sensor, unit measurements, manufacturer) from the sensor itself. Then, MOSDEN registers the newly detected sensor in the cloud GSN instance. Different MOSDEN instances  register their own sensors independently ni the cloud GSN instance. Cloud GSN combines all the information and model the data using the Semantic Sensor network ontology (www.w3.org/2005/Incubator/ssn/ssnx/ssn). 

When the cloud GSN instance receives a request from a user, it queries the sensor description registry in order to find out the relevant sensors that matches the user requirements. Then, it finds the MOSDEN instances that are capable of fulfilling the user request (i.e. whether the given MOSDEN is capable of collecting data from a sensor which is required by the user). Subsequently, GSN instance sends the requests to MOSDEN instances. Then, each MOSDEN registers the request. Finally, MOSDEN starts streaming the requested data  to the cloud GSN instance. The cloud GSN instance can make the requests in both pull and push mechanisms. In the pull method, GSN makes the request every time it wants data from MOSDEN. In push method, cloud GSN sends the request and MOSDEN sends the data back until the request expires.

\subsection{Distributed Processing}
\label{sec:A:Distributed_Processing}

Our proposed IoT middleware platform capable of running on different resource constrained mobile devices supports distribute processing. Even though in this paper, we do not discuss distributed processing in detail, it is important to note that, processing data locally saves substantial amount of network communication cost. Additionally, peer to peer data communication and processing is also important. For example, multiple MOSDEN instance can interact in peer to peer communication mode without having central controller such as cloud MOSDEN. All the processes discussed earlier is also valid in such scenarios.

\section{Implementation}
\label{sec:Implementation}

In this section, we describe the implementation details of MOSDEN. First, we present an overview of the development platforms, tools and technologies we used to develop the proposed solution. Further, we illustrate some of user interfaces provided in MOSDEN. We also discuss how we implemented the plugin architecture and the steps and guidelines that need to be followed in order to develop new plugins that are compatible with MOSDEN.

\begin{figure}[t]
 \centering
 \includegraphics[scale=1.32]{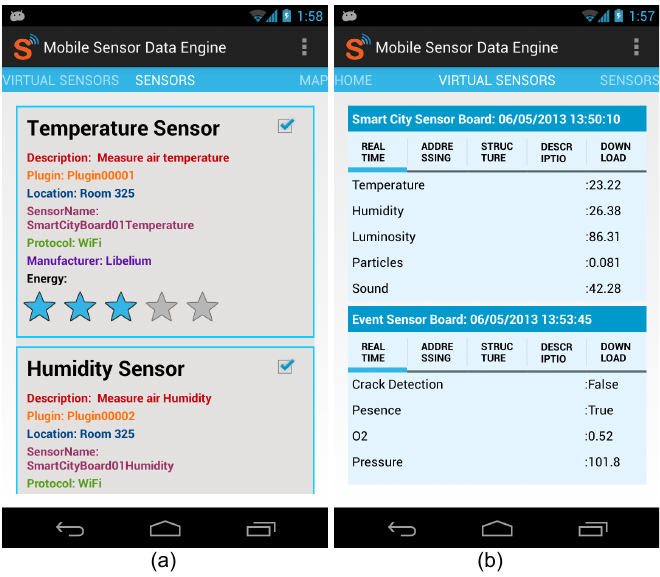}
\vspace{-0.5cm}	
 \caption{MOSDEN screenshots: (a) List of sensors connected the MOSDEN; (b) List of virtual sensors  currently running on the MOSDEN and their details}
 \label{Figure:User_Interfaces}	
\vspace{-0.93cm}	
\end{figure}

\begin{figure}[b]
 \centering
 \vspace{-0.63cm}
 \includegraphics[scale=0.88]{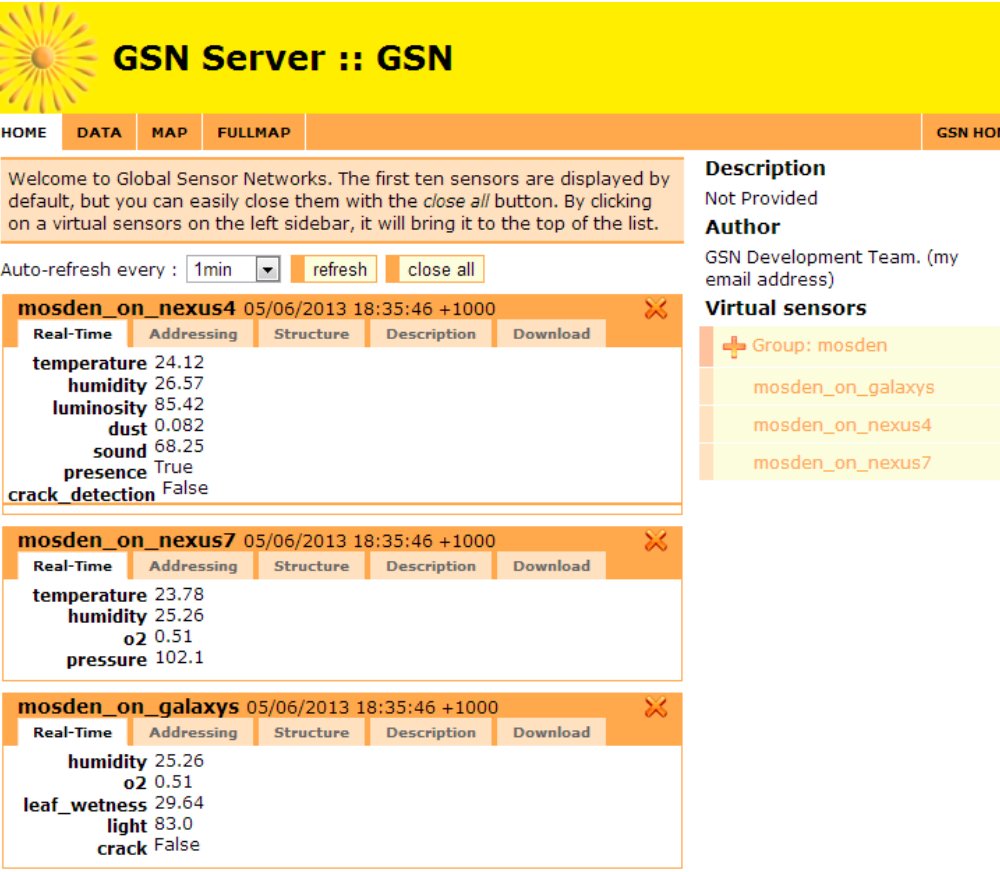}
 \caption{Screenshot of a cloud GSN instance showing three different MOSDEN instances registration}
 \label{Figure:Cloud}	
\end{figure}

Our middleware is written in Java and runs on Android based devices. We used Java to develop our middleware in order to make sure  the compatibility with its cloud based companion, GSN middleware \cite{P022}. Further, we selected Android platform due to its availability and the popularity\footnote{http://www.gartner.com/newsroom/id/2335616}. Another important factor is the portability of the Android platform. Android is not intended to be a platform only for mobile phones. The leading developer of the Andorid platform, Google Inc., intends to use if for many other smart devices such as automobiles, refrigerators, televisions, and so on. This vision supports our objectives we discussed earlier in Section \ref{sec:Background_and_Motivation}. Therefore, the objective of developing MOSDEN is not only to support mobile phone platforms but also to support devices such as \textit{Raspberry Pi} (raspberrypi.org). Currently,  android for \textit{Raspberry Pi} is under development. MOSDEN runs on Android 2.3 (and up), and it has 9935 (+ 768 logging lines in debugging version) lines of Java code. It consists of 115 class distributed across 14 packages. MOSDEN is based on popular middleware called Global Sensor Networks (GSN) \cite{P022}. MOSDEN source code will be available to  downloaded freely in the future. As we mentioned earlier, our goal is not only to support  mobile phones but also to support devices with similar resources limitations. These devices may or may not have screens. We decided to develop two different versions of our middleware based on the same underline code-base where one version provides fully-fledged user interface to support direct user interaction with the MOSDEN as illustrated in Figure \ref{Figure:User_Interfaces}. The other version provides a simple user interface that only allows to start the middleware\footnote{It is important to note that graphical user interface version requires Android 4.0 or higher as we have utilized latest user interface components in order to provide rich experience to the users. Limited user interface version is suitable for devices such as \textit{Raspberry Pi} which reduces the additional overhead caused by the user interfaces.}. Figure \ref{Figure:Cloud} illustrates the user interface of the cloud GSN.

All the features available in GSN  are also available in MOSDEN including data processing and REST-base peer-to-peer communication over HTTP. In comparison to GSN, we changed the wrapper structure and developed a generic wrapper. Further, we introduced the notion of plugins and added a plugin layer as well as  a plugin manager. We also replaced the web-based user interface with an  native Android application.

\begin{figure}[b]
 \centering
 \vspace{-0.33cm}
 \includegraphics[scale=1.1]{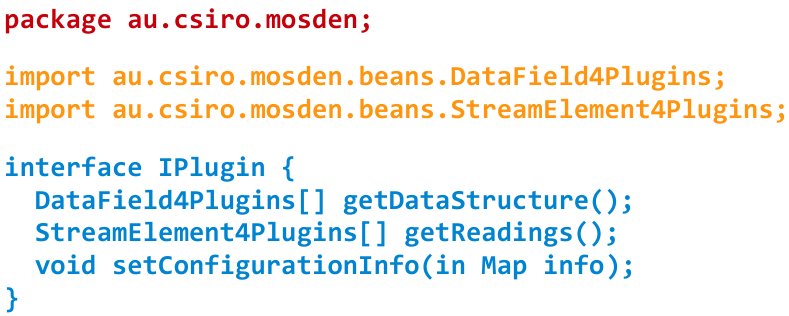}
\vspace{-0.5cm}	
 \caption{IPlugin written in AIDL (Android Interface Definition Language) that governs the structure of the Plugins. It defines the essential items in the plugin.}
 \label{Figure:AIDL_Code}	
\end{figure}

\subsection{Plugin Development}
\label{sec:M:Plugin_Development}

\begin{figure}[t]
 \centering
 \includegraphics[scale=0.9]{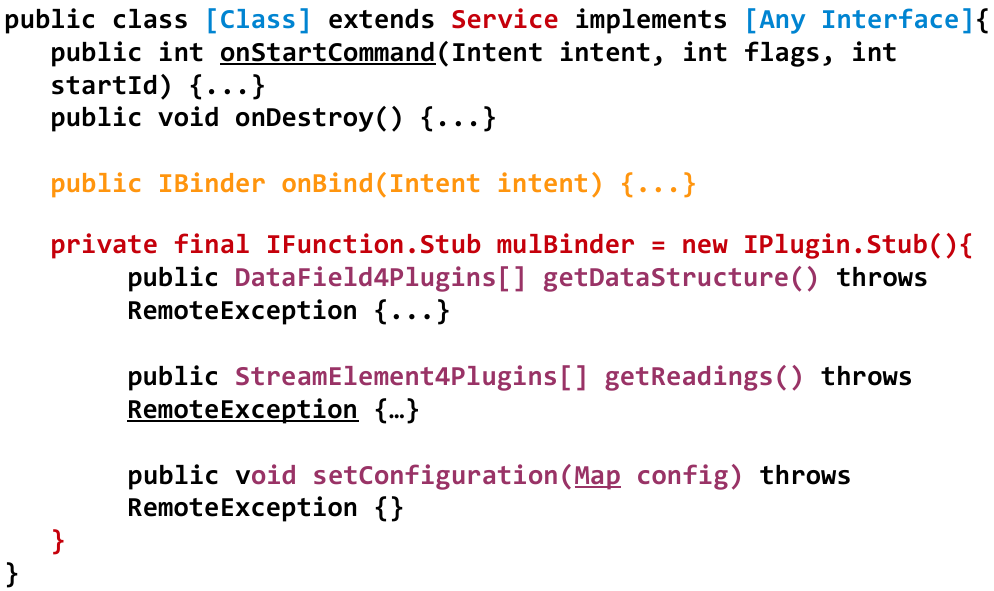}
\vspace{-0.6cm}	
 \caption{MOSDEN plugin is an Android service}
 \label{Figure:Plugin_Code}	
\vspace{-0.63cm}	
\end{figure}

This section explains how third party developers can develop plugins in such a way that their plugins  are compatible with MOSDEN so MOSDEN can use them to communicate with external sensors. In plugin development, there are three main components that needs to be considered: (1) Plugin interface written in Android Interface Definition Language (AIDL)\footnote{http://developer.android.com/guide/components/aidl.html}, (2) Plugin class written in Java, and (3) Plugin definition in \textit{AndroidManifest} file. Figure \ref{Figure:AIDL_Code} shows the plugin interface written in AIDL. \textit{IPlugin} is an interface defined in AIDL. Plugin developers should not make any changes in this file. Instead they can use this file to understand how MOSDEN plugin architecture works. \textit{IPlugin} is similar to the Java interfaces. It defines all the methods that need to be implemented by all the plugins despite their functionalities. Related to MOSDEN, we defined three methods to support the communication between main application and third party plugins\footnote{We expect to add more methods in order to support sophisticate functionalities and features in the future.}. Figure \ref{Figure:Plugin_Code} present the basic structure of a MOSDEN plugin. Each plugin is defined as an Android service. MOSDEN plugin developers need to implement these two methods: \textit{getdataStructure()} and \textit{getReadings()}. There is a third method, \textit{void setConfiguration(in Map config)}, that developers can use to retrieve data from MOSDEN at runtime, specially information unknown to them at the development time (e.g. ip address, port number and other information related to configuration). This method accepts a Map\footnote{A Java Data structure} data structure as input and does not return any output.

\begin{figure}[h]
 \centering
 \includegraphics[scale=1.02]{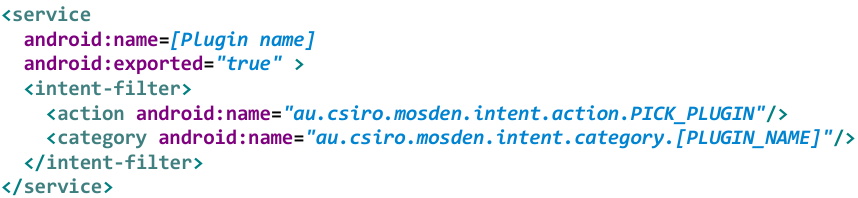}
\vspace{-0.5cm}	
 \caption{Code snippet of the plugins \textit{AndroidManifest} file}
 \label{Figure:AndroidManifest}	
\vspace{-0.33cm}	
\end{figure}

In high-level, \textit{getdataStructure()} returns a data type called \textit{DataField4Plugins[]}. This returning data structure describes what kind of data items that MOSDEN should expect from the plugin. So MOSDEN can prepare its internal data structures as necessary. At the initialization phase, MOSDEN calls the \textit{getdataStructure()}  method so MOSDEN knows to expect before real data comes in. Once the initialization is done, MOSDEN calls \textit{getReadings()} repeatedly depending on the frequency specified by the cloud GSN. The method \textit{getReadings()} returns a data raw (that comprise data items) that is organized as specified in the \textit{DataField4Plugins[]}. The return data type is \textit{StreamElement4Plugins[]}. Plugin developers are allowed to perform any operation within this method as long as it produces and returns the data types as specified by the guidelines\footnote{We expect to release a developer guide that explains how third party plugins can be developed in the future}. Figure \ref{Figure:AndroidManifest} shows how the plugins need to be defined in the AndroidManifest so MOSDEN application can automatically queried and identified them. The Android plugin must have an intent filter and the action name must be \textit{`au.csiro.mosden.intent.action.PICK\_PLUGIN'}. Developers can provide any category name based on their preferences.

\begin{figure}[h]
 \centering
 \includegraphics[scale=1.02]{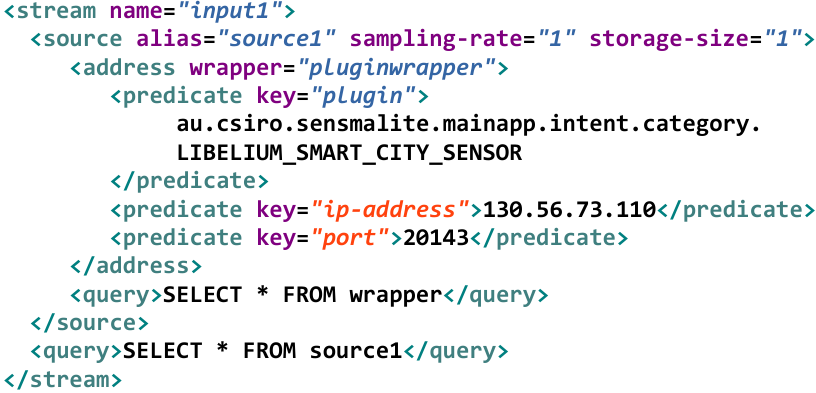}
 \caption{Code snippet of a virtual sensors definition}
 \label{Figure:VSD}	
\vspace{-0.63cm}	
\end{figure}

In order to support much user friendly and scalable plugin architecture, we extended the typical GSN Virtual Sensor Definition (VSD). The essential details that are required to connect a specific sensor to MOSDEN (e.g. IP address, port number) can be passed into the plugin via the VSD as illustrated in Figure \ref{Figure:VSD}. These details are important special, in scenarios where multiple sensors need to use the same plugin (e.g. connecting 2 sensors that are similar)

\section{Evaluation}
\label{sec:Evaluation}
In this section, we present the details of the test-beds and  evaluation methodology. We also discuss the lessons learnt from experimental evaluations.

\subsection{Test-beds}
\label{sec:M:Test_beds}

We  evaluated the proposed middleware solution, MOSDEN using several different parameters such as CPU consumption, scalability, memory requirements, latency and so on. For the evaluation, we used three devices with different resource limitations. From here onwards we refer them as D1, D2, and D3. The technical specifications of the devices are as follows.

\begin{itemize}
\item \textbf{Device 1 (D1):} Google Nexus 4 mobile phone, Qualcomm Snapdragon S4 Pro CPU, 2 GB RAM, 16GB storage, Android 4.2.2 (Jelly Bean) 
\item \textbf{Device 2 (D2):} Google Nexus 7 tablet, NVIDIA Tegra 3 quad-core processor, 1 GB RAM, 16GB storage, Android 4.2.2 (Jelly Bean)
\item \textbf{Device 3 (D3):} Samsung I9000 Galaxy S, 1 GHz Cortex-A8 CPU, 512 MB RAM, 16GB storage, Android 2.3.6 (Gingerbread)
\end{itemize} 

\begin{figure}[h]
 \centering
 \includegraphics[scale=0.45]{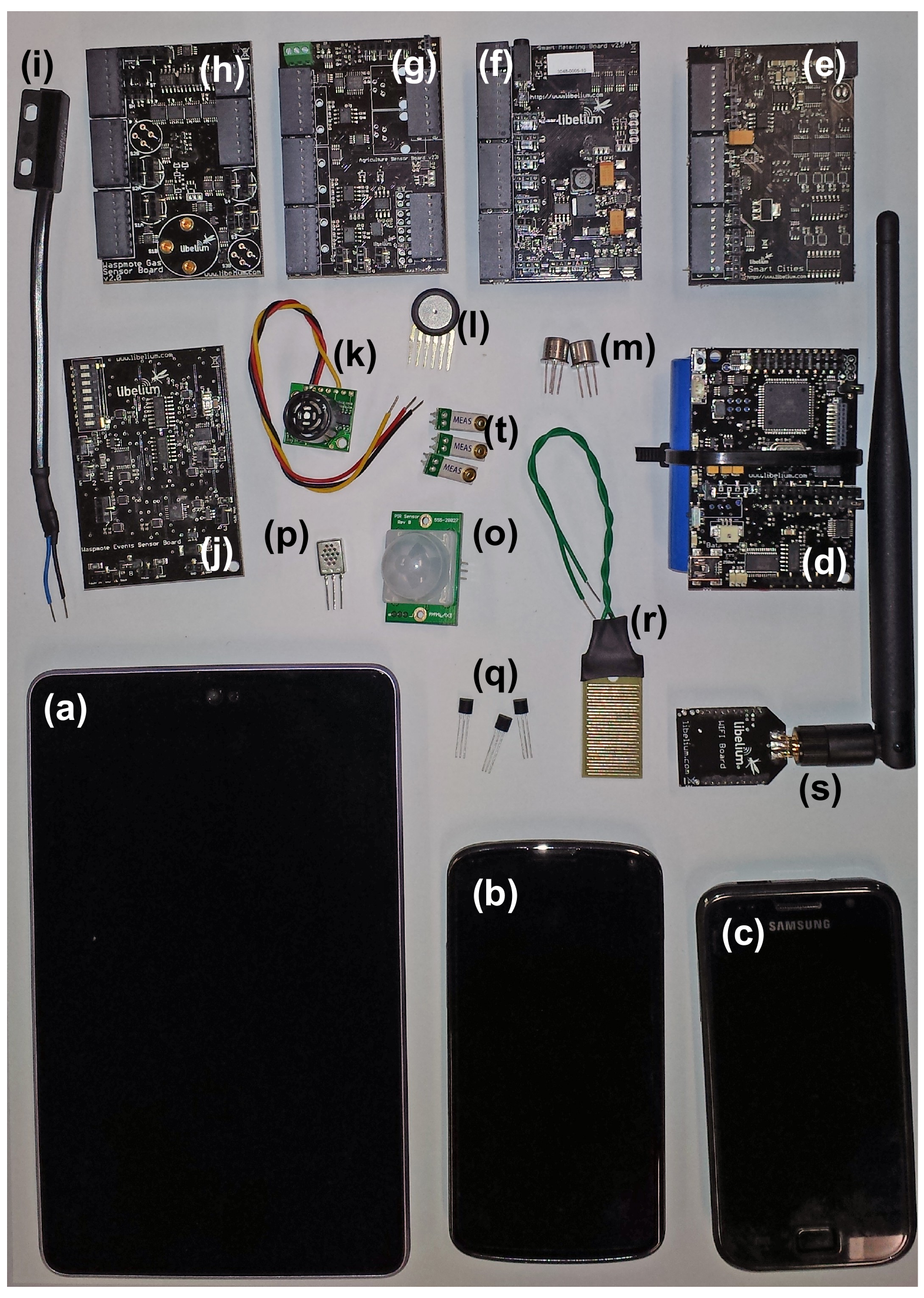}
 \caption{Some of the hardware devices used in the experimentation: (a)  Google Nexus 7 tablet, (b) Google Nexus 4 mobile phone, (c) Samsung I9000 Galaxy S, (d) Smart cities board (with batteries), (e) Smart cities board (without batteries), (f) Smart metering board, (g) Agriculture board, (h) Gas board, (i) Flex bend sensor, (j) Events board, (k) Ultra sound sensor ,(l) Air pressure sensor,(m) Air contaminant sensors, (o) Presence sensor,(p) Humidity sensor,(q) Temperature sensors, (r) Leaf wetness sensor, (s) Wi-Fi broad / Antenna, (t) Vibration sensors}
 \label{Figure:Hardware}	
\vspace{-0.83cm}	
\end{figure}

We used a computer with Intel(R) Core i5-2557M 1.70GHz CPU  and 4GB RAM to  host the cloud GSN during the evaluations. For our evaluations, we employed sensors built into the above devices (e.g. Motion sensors: accelerometer, gravity, gyroscope, liner acceleration, rotation vector;  Environmental sensors: ambient temperature, light, pressure, relative humidity; Position sensors: magnetic fields, orientation, proximity.). Further, we used sensors manufactured by Libelium \cite{P595} as external sensors with different combination of hardware sensors plugged into them such as    temperature sensor, humidity sensor, LDR sensor, air pressure sensor, leaf wetness sensor, noise sensor, dust sensor, force and pressure sensor, flex-bend sensor, flexible stretch sensor, hall-effect sensor, differnt gas sensors (e.g. O$_{2}$, CO$_{2}$) and so on. Resource constrained computational devices we used in this work as well as some of the sensors used in this experiments are shown in Figure \ref{Figure:Hardware}.

\begin{figure*}[t!]

        \centering
        \begin{subfigure}[b]{165pt}
                \centering
                \includegraphics[scale=0.38]{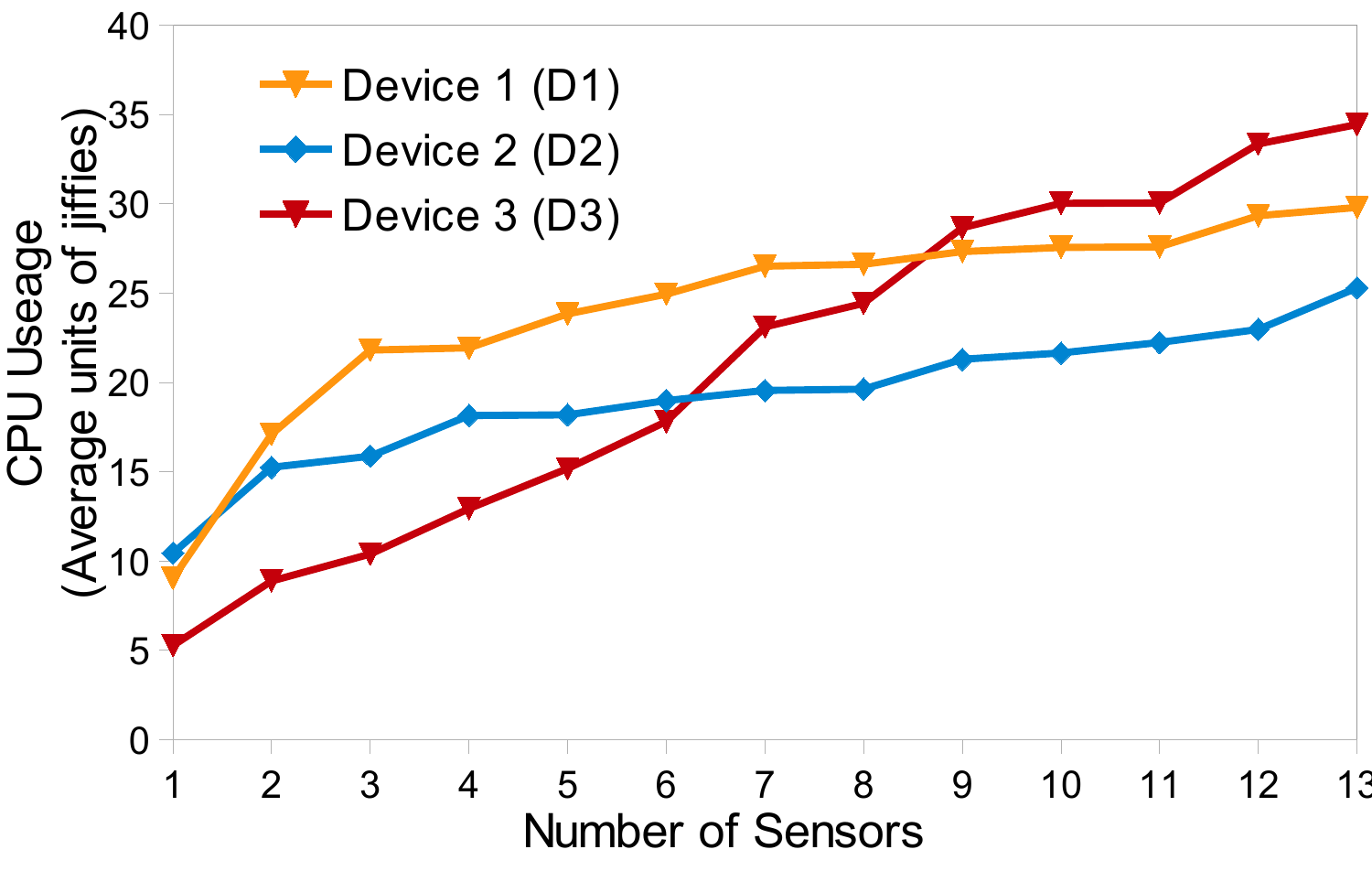}
                \vspace{-16pt}
                \caption{\footnotesize }
                \label{Figure:Results1}
                \vspace{-8pt}
        \end{subfigure}%
        ~ 
        \begin{subfigure}[b]{165pt}
                \centering
                \includegraphics[scale=.38]{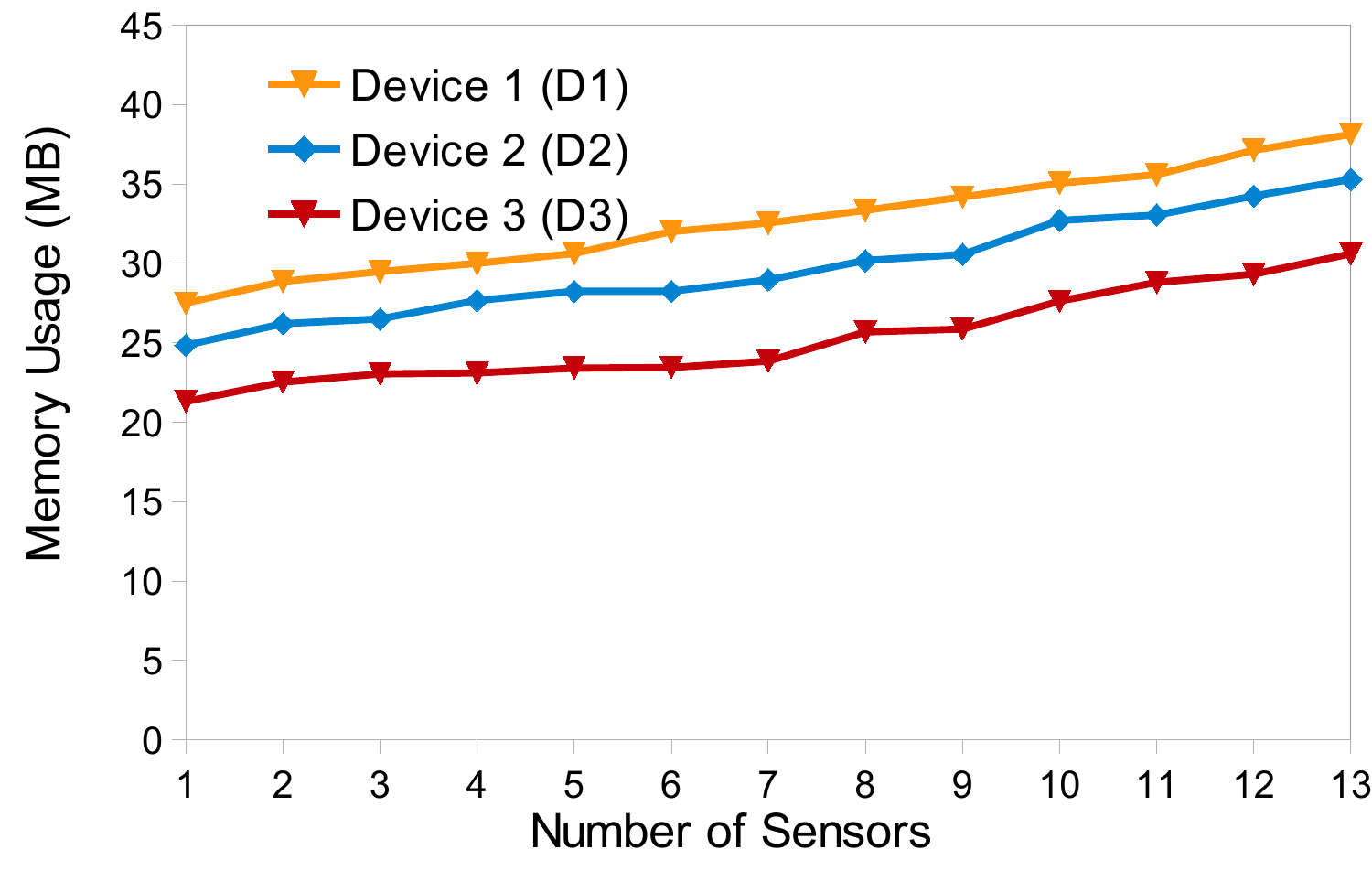}
                \vspace{-16pt}
                \caption{\footnotesize }
                \label{Figure:Results2}
                \vspace{-8pt}
        \end{subfigure}
        ~ 
        \begin{subfigure}[b]{165pt}
                \centering
                \includegraphics[scale=.38]{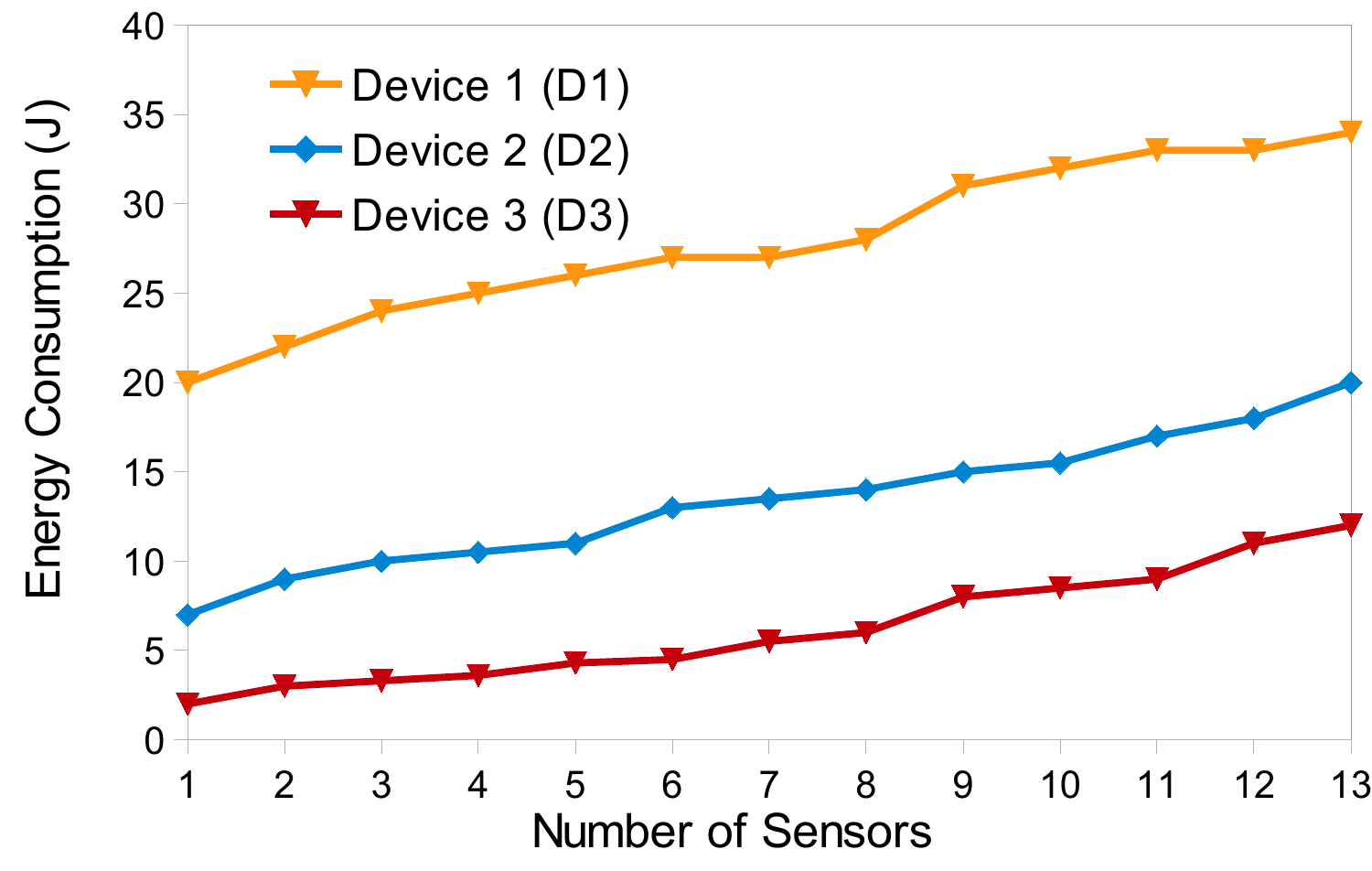}
                \vspace{-16pt}
                \caption{\footnotesize }
                \label{Figure:Results3}
                \vspace{-8pt}
        \end{subfigure}

        \begin{subfigure}[b]{165pt}
                \centering
                \includegraphics[scale=0.38]{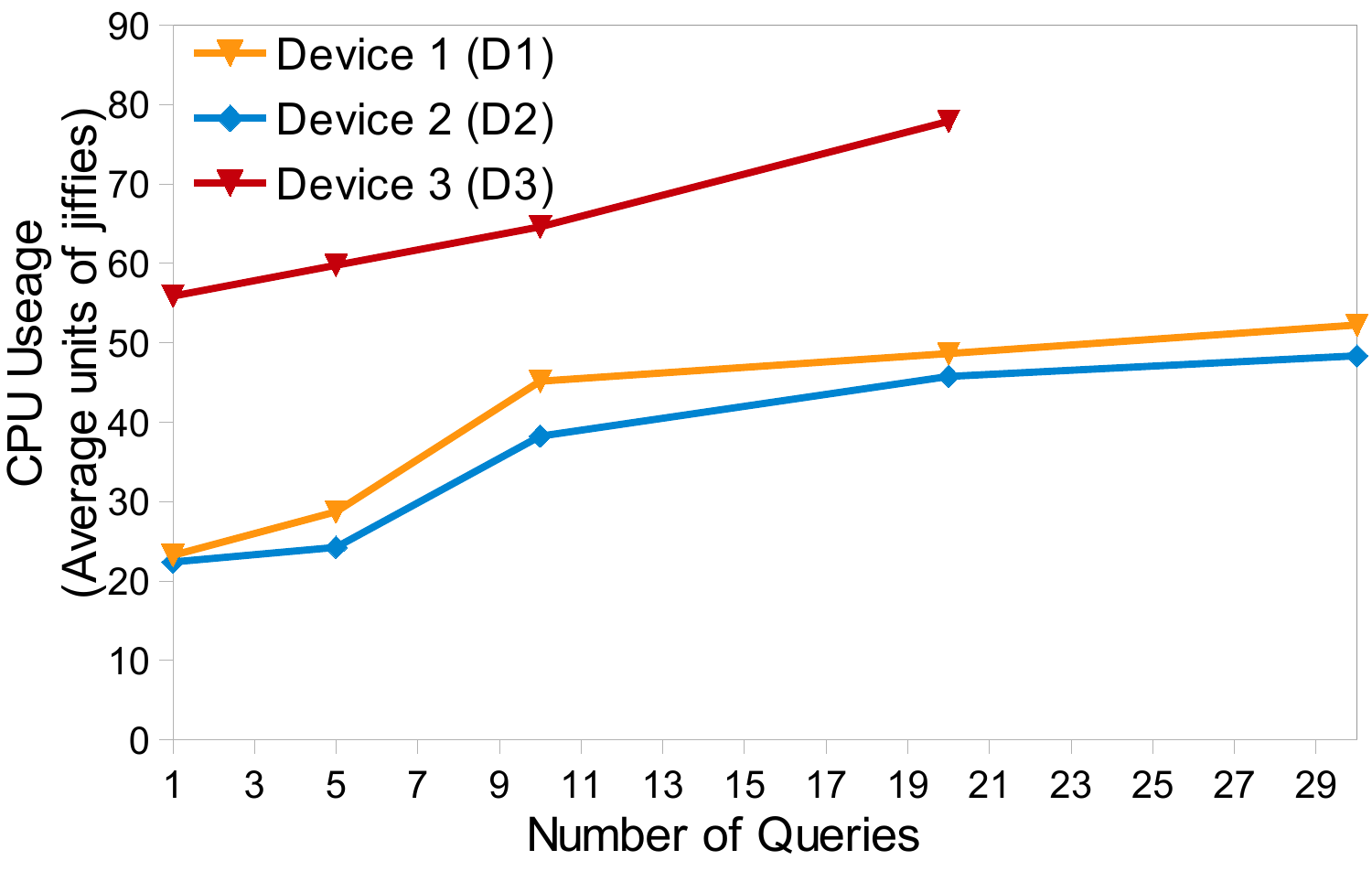}
                \vspace{-16pt}
                \caption{\footnotesize }
                \label{Figure:Results4}
        \end{subfigure}%
        ~ 
        \begin{subfigure}[b]{165pt}
                \centering
                \includegraphics[scale=.38]{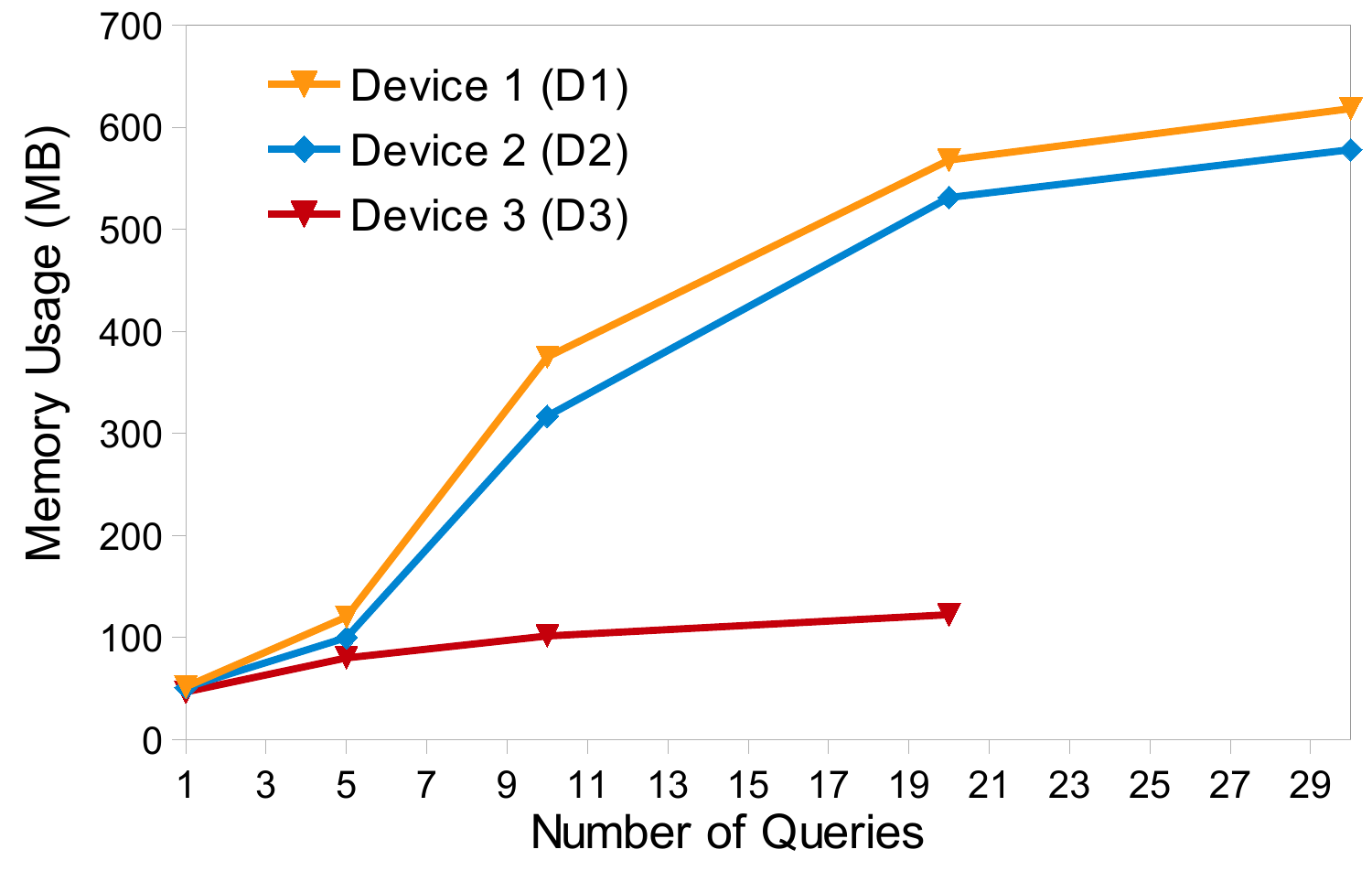}
                \vspace{-16pt}
                \caption{\footnotesize }
                \label{Figure:Results5}
        \end{subfigure}
        ~ 
        \begin{subfigure}[b]{165pt}
                \centering
                \includegraphics[scale=.38]{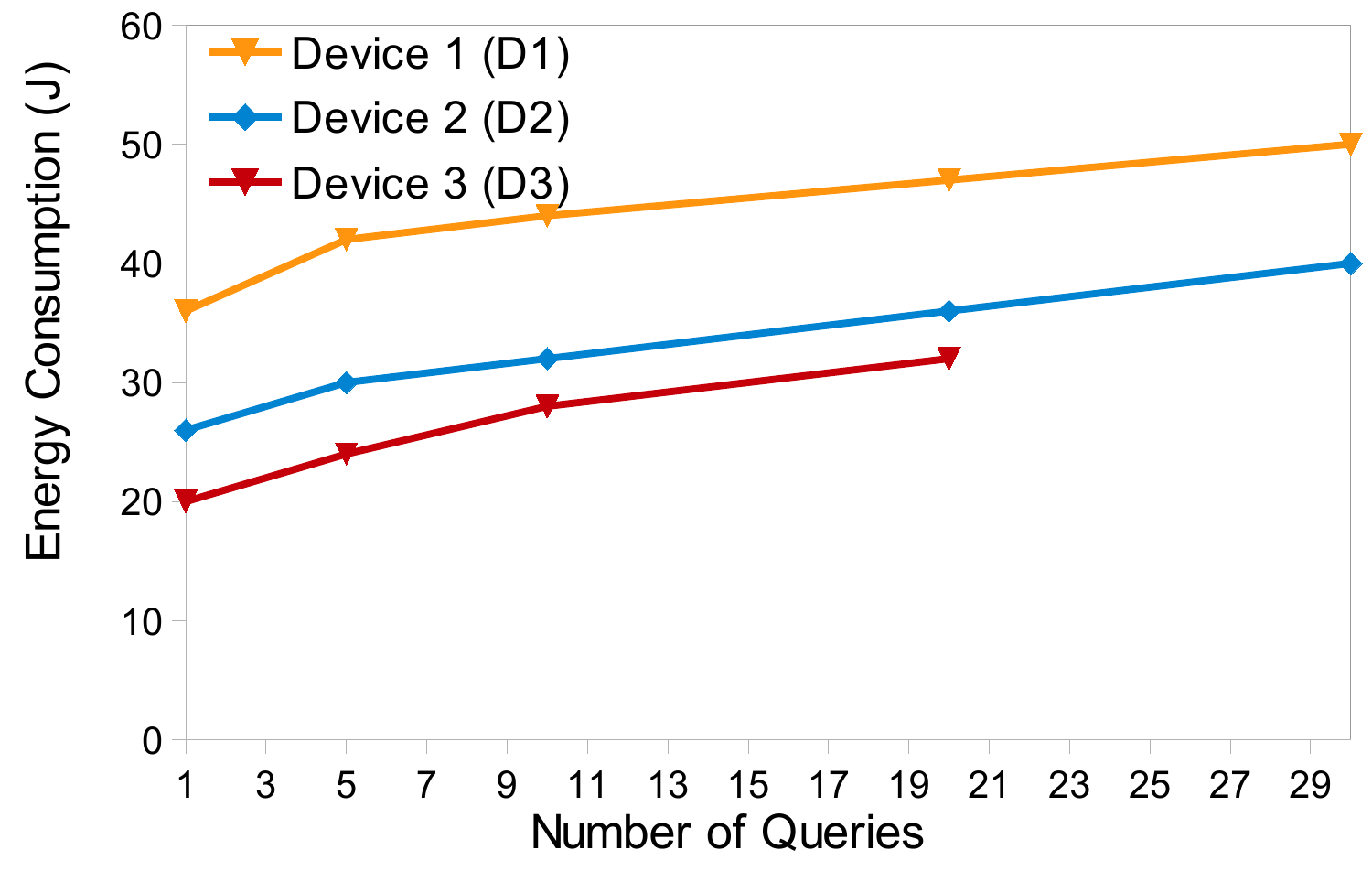}
                \vspace{-16pt}
                \caption{\footnotesize }
                \label{Figure:Results6}
        \end{subfigure}
        \vspace{-10pt}
        \caption{Experimentation and Evaluation Results. (Note: MOSDEN application and plugins use CPU, memory, and energy independently where plugins are treated as individual service by Android platform. All the calculations are accumulated values of MOSDEN application and plugin services)}
        \label{Figure:Resulsts}
        \vspace{-20pt}
\end{figure*}

\subsection{Methodology}
\label{sec:M:Methodology}

\begin{figure}[b]
 \centering
 \vspace{-10pt}
\includegraphics[scale=0.45]{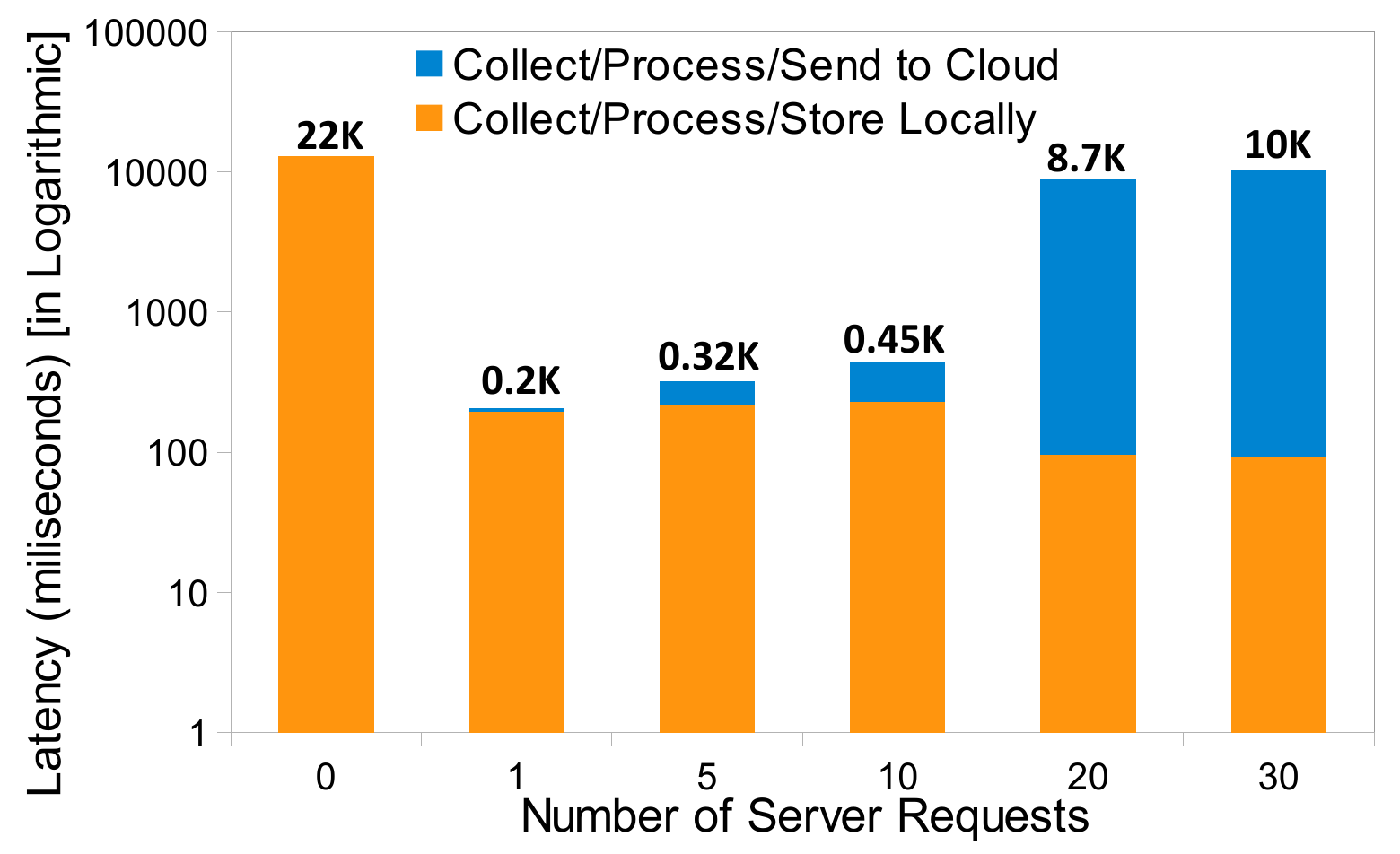}
\vspace{-8pt}
\caption{Data Processing Latency}
\label{Figure:Results7}	
\end{figure}

This section explains the evaluation methodology, experimental conditions and objectives of the Figures \ref{Figure:Results1} to \ref{Figure:Results7}. All the evaluations are done using three different resource constrained mobile devices as explained in the section above. In all the evaluations, CPU usage (consumption) is measured in units of \textit{jiffies}\footnote{In computing, a jiffy is the duration of one tick of the system timer interrupt. It is not an absolute time interval unit, since its duration depends on the clock interrupt frequency of the particular hardware platform.}. At this point, MOSDEN supports only Wi-Fi communications\footnote{We expect to support Zigbee and Bluetooth in the future. However, such improvements will not change the overall architecture.}. We keep the sampling rate as 1 second during the course of evaluations.

In Figure \ref{Figure:Results1}, we examine how CPU usage changes when the number of sensors involved increases. Figure \ref{Figure:Results2} shows how memory consumption changes when number of sensors involved increases. Figure \ref{Figure:Results3} measures how energy consumption  changes when number of sensors involved increases. In Figures \ref{Figure:Results1}, \ref{Figure:Results2} and \ref{Figure:Results3}, MOSDEN only uses  inbuilt sensors to collect data and store them in the local storage space. No network communication is performed. In Figure \ref{Figure:Results4}, we evaluate how CPU usage changes when the  number of queries processed by the MOSDEN changes (step 2 and 3 in Figure \ref{Figure:System_Overview}). Figure \ref{Figure:Results5} shows how memory consumption changes when the number of queries changes. Additionally, Figure \ref{Figure:Results6} shows how energy consumption  changes when the number of queries changes. In Figures \ref{Figure:Results4}, \ref{Figure:Results5}, and \ref{Figure:Results6}, MOSDEN uses inbuilt sensors to collect data and send them to the cloud GSN over a WiFi network.

In Figure \ref{Figure:Results7}, we examine the time MOSDEN takes (i.e. latency) to process and transmit the data. We measure the time taken for the following two operations. (1) We start measuring the time taken by the plugin to retrieve data from a sensor, pass it to a wrapper and subsequently store it in a local database. (2) Time taken for MOSDEN to respond to incoming query request from the cloud GSN.


\subsection{Results}
\label{sec:M:Results}

According to Figure \ref{Figure:Results1}, it is evident that CPU usage increases when the number of sensors increase. It is important to highlight that, D3 consumes more CPU time compared to other devices when it needs to handle 10+ sensors. One reason for this is the lack of main memory (RAM) which puts additional overheads on the CPU. Similar pattern is revealed in Figure \ref{Figure:Results2} as well as in term of memory usage. Devices that have larger memory capacity can afford to allocate more memory to MOSDEN which increases the overall performance of MODSEN. Further, comparatively resource rich devices consumes more energy due to usage of powerful CPUs and sensing hardware. This is observed in the \ref{Figure:Results3} where difference in energy consumption for D1 and D3 is much higher compared to difference in memory usage. When not performing any network communication tasks, MOSDEN takes only 38MB (D1) / 30MB (D3) to collect, process and store data from 13 different sensors\footnote{All the devices do not have all 13 sensors though the Android platform supports them}. MOSDEN consumes around 35J (D1) / 10J (D3) to process, and store data from 13 sensors. It is important to note that, Android manages the memory allocated to application. Depending on the memory availability at a given point of time, Android could restrict an application from consuming large amount of memory to facilitate smooth running of other essential applications and services. We also note, during this evaluation, only system processes and services, MOSDEN and our power monitor application were running on the phone.


Most important fact is that D3 could not handle more than 20 parallel queries from the cloud GSN\footnote{When running more than 20 queries in D3, MOSDEN becomes unstable. Sometimes, D3 was able to handle 20+ queries for a very short period of time before it crashed}. This is mainly due to lack of available memory as compared to other devices. Additionally, D3 is based on Android 2.3.6 (Gingerbread) OS and does not support multi core operations. Further, Android 2.3.6 does not provide efficient multi tasking support\footnote{http://socialcompare.com/en/comparison/android-versions-comparison}. As we mentioned earlier related to Figure \ref{Figure:Results1}, Figure \ref{Figure:Results4} also reveals that D3 uses significantly more CPU compared to other devices due to the overhead created by lack of memory. Comparatively, D1 and D2 use less CPU and as observed from the results, the CPU consumption is gradually increasing but not significant when MOSDEN processes more than 10 queries. One reason for this is that, Android OS restricts MOSDEN from consuming too much CPU resource after a certain level as it needs to facilitate other essential Android applications and services.

Figure \ref{Figure:Results5} clearly shows that D3 suffers from lack of memory as it is not allocated more than 150MB of memory. In contrast, both D1 and D2 have abundant memory available to be utilized so memory usage increase up to 620MB (D1) / 580MB (D2). Energy consumption graph with and without network communication looks similar in pattern. However, energy consumption has significantly increased across all three devices (50J (D1) / 40J (D2) when processing 30 queries). When there is no network communication, MOSDEN takes 22 seconds to collect data from a sensor plugin, process and store it locally. However, when the cloud GSN starts sending queries Android allocates more CPU and memory to MOSDEN. Hence, the data collection/processing and query processing operations are performed much faster which helps to reduce the overall latency from 22 seconds to 0.2 seconds. As the number of query request increases, from the results, we observe that, latency\footnote{Time takes to fulfil all the requests made by the cloud GSN} also increases. When MOSDEN processes 30 queries, latency increases to 10 seconds. However, significant potion of the total processing time is taken to fuse the data and send them to the cloud\footnote{Time that the data takes to travel over the network is not counted}.

%
%
%
%
%
%
%
%
%

\subsection{Lessons Learnt and Potential Applications}
\label{sec:M:Discussion_and_Lessons_Learnt}

\textbf{Lessons Learnt:} Our experimental evaluations validate the energy and performance efficient of our the proposed plugin-based MOSDEN platform. The middleware functioned without any issues during our experiments. Additionally, the plugin-based architecture increases the usability of MOSDEN by allowing users to download and install plugins from Google market place with zero effort programming and no modifications to MOSDEN. Further, modern mobile devices can process significant amount of requests with the limited resources they have. It is evident that the memory is more important than CPU in situation where data needs to be processed under small sampling rates. In our previous work \cite{ZMP001}, we learnt that reduced sampling rate can  save energy and resource consumption significantly. In such scenarios, MOSDEN will be able to process much more queries efficiently than it did in the evaluations. We look forward to perform more experiments to examine the impact of sampling rate on MOSDEN's performance.

\textbf{Potential Applications:} The MOSDEN platform can greatly fosters the development of new and innovative mobile data services that depend on IoT devices as the source of data. One such example is a crowd-sourcing application where sensor data (e.g. noise level in outdoor environment) can be collected from users' mobile device running MOSDEN. The collected data can be used by applications in the cloud in their decision making process (e.g. determine the noise pollution level at an intersection in the city by fusing data from multiple MOSDEN instances). Another example is to determine real-time traffic conditions using data acquired from MOSDEN running on user mobile devices.


\section{Related Work}
\label{sec:Related_Work}
In this section, we review the literature under three categories of research  where our proposed solution lies at the intersection: IoT middleware, mobile Sensor middleware, and data processing in resource constrained devices.

\textbf{IoT Middleware:} Bandyopadhyay et al. \cite{P118} have done a survey on IoT middleware solutions. We discussed the GSN middleware in detail before as we selected it to build our solution on top of it.  Microsoft \textit{SensorMap} \cite{P578} (sensormap.org) is a data sharing and visualization framework. It is also a peer produced sensor network that consists of sensors deployed by contributors around the world. \textit{SensorMap} mashes up sensor data on a map interface. Then, it allows to selectively query sensors and visualize data. \textit{SensorMap} is designed to run on server computers. \textit{Linked Sensor Middleware} (LSM) \cite{P584} (lsm.deri.ie) is a platform that provides wrappers for real time data collection and publishing. It also provides a web interface for sensor search, linked stream data query, data annotation and visualisation. \textit{LSM} mainly focuses on linked data publishing. This platform is also focused on server-based deployment. \textit{Cosm} (formerly Pachube) (cosm.com) is a platform for Internet of Things devices. \textit{Cosm} allows different data sources to be connected to it. Then, it provides functionalities such as event triggering and data filtering. \textit{Cosm} is also a server level middleware that is not suitable for resource constraint devices. There are several other commercial solutions available: TWINE (supermechanical.com), Ninja Blocks (ninjablocks.com), and Smart Things (smartthings.com). All these solutions focus on event detection using IF-THEN rules in smart environments. They use their own proprietary software systems installed on small resource constrained computational devices. Their own sensors can communicate with these devices. MoSHub \cite{ZMP005} is an mobile application that collects data from both internal and external sensors and push them to the cloud IoT middleware. However, it does not provide data processing capabilities. This means all the sensor data collected are uploaded to the cloud then and there which makes this approach inefficient in term of energy consumption.

\textbf{Mobile Sensor Middleware:} These category of solutions aim  to turn smart phones into mobile sensors so data on user behaviour and the surrounding environment  can be captured and analysed. Mobile phone based sensing algorithms, approaches, and applications are discussed in \cite{P217}.  \textit{Pogo} \cite{P628} is a middleware for mobile phone sensing that focuses on building large scale mobile phone sensing test beds. They have developed a server-based counterpart as well as the middelware for Android. The objective of \textit{pogo} is to sense the behaviour of the applications. In contrast, MOSDEN focuses on collecting and processing sensor data from both internal and external sensors. DAM4GSN \cite{ZMP001} is also an approach based on GSN. It provides an application that is capable of collecting data from internal sensors of a mobile phone and send to the GSN middleware. No processing capabilities are provided at the mobile phone end. Therefore, all the information sensed are sent to the server. This approach is inefficient due to continuous usage of communication radio of the mobile phone specially when the sampling rate is small ($<$ 1 min) \cite{ZMP001}.

\textbf{Data Processing in Resource Constrained Devices:} \textit{Dynamix} \cite{P627} is a plug-and-play context framework for Android. \textit{Dynamix} automatically discovers, downloads and installs the plug-ins needed for a given context sensing task. \textit{Dynamix} is a stand alone application and it tries to understand new environments by using pluggable context discovery and reasoning mechanisms. It does not provide server-level solution. Context discovery is the main functionality in \textit{Dynamix}. In contrast, our solution is focused on allowing easy way to connect sensors to applications in order to support sensing as a service model in IoT domain. We employee a pluggable architecture  which is similar to the approach used in \textit{Dynamix}, in order to increase the scalability and rapid extension development by 3rd party developers. One of the most popular type of processing in mobile is activity recognition. Yan et al. \cite{P629} have presented an energy-efficient continuous activity recognition on mobile phones. One of the most important data processing task that need to be performed at the lower level devices (e.g. categories 2,3,4 in Figure \ref{Figure:Layered_Architecture}) in IoT is validation, fusing, filtering, context discovery, and annotation \cite{ZMP007}. Data collected by lower level devices needs to be validated in order to reduce the wastage of network communication \cite{P630}. Further, data fusing and filtering operations prevent redundant  network communications. Some of the context information such as location need to be discovered at the lowest layers \cite{P631} which makes it impossible to perform such operation at higher levels.

\section{Conclusion and Future Work}
\label{sec:Conclusion}
The number of mobile devices connected to the Internet is growing at a rapid space. Significant portion of these devices are mobile phones today. However, it is expected that billions of different types of resource constrained computational device will be connected to the Internet over the coming decade. On the other hand, number of sensors deployed around us are getting increased. It is an increasingly important task to collect data from these sensors in order to analyse and act upon them.

In this paper, we presented an Internet of Things middleware for resource constrained computational mobile devices called MOSDEN. Our proposed middleware also supports sensing as a service model. MOSDEN provides an easy and convenient way to connect sensors to the mobile device with zero programming effort. We introduced a scalable plugin architecture where plugins are distributed through leading mobile application stores such as \textit{Google Play}. MOSDEN is capable of collecting data from multiple different sensors and process them together. MOSDEN is 100\% compatible with Global Sensor Network Middleware that runs on the cloud. Further, MOSDEN can act as a peer to peer data processing engine as well. We evaluated MOSDEN in different aspects such as resource consumption, scalability, and usability. We have demonstrated the feasibility and scalability towards using MOSDEN on resource constrained devices to collect and process sensors data. We also demonstrated that significant amount of resources can be saved by processing the data locally instead of transmitting all data to the remote server. It is evident that such processing allows to run the IoT infrastructure for longer time period. 

\textcolor{black}{We will release the source code of MOSDEN platform to the public in future.} In the future, we would like to add automated sensor discovery and configuration functionalities to the MOSDEN where it will be able to search and discover any kind of sensors around a given location and automatically install the required plugins. This will allows MOSDEN to communicate and configure the sensors autonomously. Further, we will develop a configuration model that can be used to configure different devices (belongs to different categories) in the IoT architecture  presented in Figure \ref{Figure:Layered_Architecture}. Configuration details and sensing strategies such as scheduling, sampling rate, data acquisition method, and protocols will be designed for each individual sensor by the higher-level devices and will be pushed towards the lower layers. Further, we will investigate the impact of employing different data processing techniques and sampling rates in order to find-out their the suitability towards resource constraint devices considering  the factors such as energy consumption and network communication.

\textbf{Acknowledgements:} Authors acknowledge support from OpenIoT Project, which is co-funded by the European Commission under seventh framework program FP7-ICT-2011-7-287305-OpenIoT.

\vspace{-8pt}



%
  \bibliography{Bibliography}
  \bibliographystyle{abbrv}

\end{document}